\documentstyle[prb,aps,twocolumn]{revtex}  
\input epsf
\epsfverbosetrue
\tighten
\begin{document}
\draft
\twocolumn[\hsize\textwidth\columnwidth\hsize\csname @twocolumnfalse\endcsname

\title{Statics and dynamics of domain patterns in hexagonal-orthorhombic 
ferroelastics}
 
\author{S. H. Curnoe\cite{Stephanie} and A. E. Jacobs\cite{Allan}}

\address{Department of Physics, University of Toronto\\
Toronto, Ontario, CANADA M5S 1A7}
 
\date{\today}
 
\maketitle
 
\begin{abstract}

We study the statics and the dynamics of domain patterns in proper 
hexagonal-orthorhombic ferroelastics; 
these patterns are of particular interest because they provide a rare 
physical realization of disclinations in crystals. 
Both our static and dynamical theories are based entirely on classical, 
nonlinear elasticity theory; 
we use the minimal theory consistent with stability, symmetry and 
ability to explain qualitatively the observed patterns. 
After scaling, the only parameters of the static theory are a temperature 
variable and a stiffness variable. 
For moderate to large stiffness, our static results show nested stars, 
unnested stars, fans and other nodes, triangular and trapezoidal regions of 
trapped hexagonal phase, {\it etc} observed in electron microscopy of Ta$_4$N 
and Mg-Cd alloys, and also in lead orthovanadate (which is 
trigonal-monoclinic); 
we even find imperfections in some nodes, like those observed. 
For small stiffness, we find patterns like those observed in the mineral 
Mg-cordierite. 
Our dynamical studies of growth and relaxation show the formation of these 
static patterns, and also transitory structures such as 12-armed bursts, 
streamers and striations which are also seen experimentally. 
The major aspects of the growth-relaxation process are quite unlike those in 
systems with conventional order parameters, for it is inherently nonlocal; 
for example, the changes from one snapshot to the next are not predictable by 
inspection. 

\end{abstract}
 
] 
\narrowtext
\tightenlines
 
\section{introduction}

Ferroelastics\cite{aizu69,salje93} are solids that undergo diffusionless, 
shape-changing phase transformations, usually first-order. 
The low-temperature (low-$T$) phase is, in principle, one of several degenerate 
products (``variants'') differing only in orientation. 
But one rarely sees the homogeneous product phase; 
instead, multiple nucleation events, constraints, intrinsic inhomogeneities, 
{\it etc}, produce domain structures composed of multiple variants separated 
by domain walls (ideally with certain orientations\cite{sapriel75}). 

This article deals with hexagonal-orthorhombic (H-O) ferroelastics, where the 
three equivalent stretch directions give rise to three variants. 
Examples are 
the mineral Mg-cordierite (Mg$_2$Al$_4$Si$_5$O$_{18}$, 
References\onlinecite{salje93,venk54,putnis87,muller91}), 
Ta$_4$N (Reference\onlinecite{vicens76}), 
and Mg-Cd alloys (References\onlinecite{sinclair77,kitano88}). 
The most interesting feature of these materials is that they provide a rare 
physical realization of disclinations in crystals\cite{sinclair77,kitano88}. 
The disclinations result because domain walls rotate the variants as well as 
link them, all nodes but one failing to close to $360^\circ$; 
the consequence is the fascinating domain patterns observed in these materials 
and in the closely related trigonal-monoclinic material lead orthovanadate 
[Pb$_3$(VO$_4$)$_2$, References\onlinecite{amelinckx80a,amelinckx80b}]. 
These patterns differ strikingly from those in cubic-tetragonal (C-T) and 
tetragonal-orthorhombic (T-O) systems; 
for example, the star disclination, often self-similarly nested, has no 
counterpart in the other systems. 

Of the several theories\cite{salje87,reid97,blackjphy99,blackpcm99,wen99} 
dealing with these materials, only that of Wen {\it et al.}\cite{wen99} 
found patterns consistent with some of those observed. 
Their phase-field kinetic theory is however highly phenomenological, being 
based on order parameters with no direct physical interpretation; 
an expansion to sixth order is necessary, and even the scaled theory requires 
at least five parameters. 

We investigate H-O ferroelastics by expanding the energy density in physical 
quantities only, namely the strains and their derivatives, to fourth order 
in the primary order parameter. 
Computational resources allow us to consider only two-dimensional 
structures, but this is not much of a restriction, for these are what the 
literature reports. 
Our density respects the symmetry of the high-$T$ phase; 
it is minimal, for it contains only essential terms. 
Scaling reduces the parameters in the density to two, a temperature variable 
and a stiffness variable, both readily determined from experiment. 
Our calculated structures depend strongly on the stiffness, which determines 
how the system responds to the disclinations generated by wall junctions. 
We investigate both the statics and the dynamics of the theory, using in both 
methods which satisfy the compatibility relations exactly. 

Our static theory, described in Section II, is based solely on classical 
elasticity theory; 
it includes nonlinear and gradient terms but no additional phenomenology. 
Section III solves for the two kinds of wall linking a given pair of variants. 
Section IV describes static structures found numerically. 
Section V first describes the equations of motion of our dynamical theory 
(also based solely on classical elasticity theory) and then goes on to 
present structures as they evolve in time. 

Although our theory is strictly valid only for proper ferroelastics, where 
the strain is the primary order parameter, we find that it is sufficient to 
explain qualitatively the patterns observed in all four materials listed 
above, namely several kinds of three-fold nodes, a common four-fold node, 
stars (both nested and unnested), the fan node and another untitled node, 
and even imperfections in some higher-order nodes. 
Near $T_c$, we find triangular and trapezoidal pockets of the high-$T$ phase, 
also as observed\cite{amelinckx80a,amelinckx80b}. 
Many of these structures were not found by Wen {\em et al.}\cite{wen99}. 
Our dynamical studies start from the perturbed high-$T$ phase. 
In the growth stage, we see 12-armed bursts as observed in Mg-cordierite. 
In the relaxation stage, but still far from the converged state, we see 
streamers and striations also as observed, and also the formation of the 
structures seen in our static studies. 
These are the first simulations to report the bursts, streamers and 
striations. 

\section{strains and strain energy}

Hexagonal-orthorhombic ferroelastics require five strains in general. 
Since we consider only columnar structures uniform in the 3 direction (normal 
to the basal plane), we need only the dilatational strain $e_1$ and 
the deviatoric/shear strains $e_2$ and $e_3$. 
In the linear approximation for the strain tensor, 
\renewcommand{\theequation}{1\alph{equation}}
\setcounter{equation}{0}
\begin{eqnarray} 
e_1&=&\left(u_{1,1}+u_{2,2}\right)/2\ ,\\ 
e_2&=&\left(u_{1,1}-u_{2,2}\right)/2\ ,\\ 
e_3&=&\left(u_{1,2}+u_{2,1}\right)/2\ , 
\end{eqnarray} 
\renewcommand{\theequation}{\arabic{equation}}
\setcounter{equation}{1}
where ${\bf u}(x_1,x_2)$ is the displacement of the material point originally 
at ${\bf x}$, and $u_{i,j}=\partial_ju_i=\partial u_i/\partial x_j$. 

The free energy $F$, per unit length in the 3 direction, is the integral 
$F=\int_A {\cal F}\ d^2x$ 
of the free-energy density ${\cal F}$ over the undeformed area $A$. 
Since the following is the first study of structures in proper H-O 
ferroelastics, and since there is no evidence that other terms are required, 
we use the simplest possible density consistent with symmetry, stability and 
ability to explain qualitatively the structures observed. 
Of course the terms omitted may be numerically important; 
in fact a class of these terms is qualitatively important, for we show below 
that it breaks an important degeneracy. 

The minimal density is slightly simplified from one used 
previously\cite{reid97}:
\begin{eqnarray} 
{\cal F}= 
& & \frac{A_1}{2}      e_1^2
+\frac{A_2}{2}\left(e_2^2+e_3^2\right)
-\frac{B_2}{3}\left(e_2^3-3e_2e_3^2\right) \nonumber\\
&+&\frac{C_2}{4}\left(e_2^2+e_3^2\right)^2
 + \frac{d_2}{2}\left[\left(\vec\nabla e_2\right)^2
                     +\left(\vec\nabla e_3\right)^2\right] \ .
\end{eqnarray} 
All coefficients but $A_2$ are positive and independent of $T$. 
The $A_1$ and $A_2$ terms are the contributions from homogeneous volume 
changes and shears respectively, the latter in the high-$T$ phase only; 
the parameters $A_1$ and $A_2$ are related to the Voigt constants by 
$A_1=2(C_{11}+C_{12})$ and $A_2=2(C_{11}-C_{12})=4C_{66}$. 
But $A_2$ softens with decreasing $T$, as $A_2=a(T-T_0)$, where $T_0$ is the 
stability limit of the H phase and $a$ is a material-dependent constant. 
The instability due to the vanishing of $A_2$ is isotropic. 
The orientations of the three O variants are selected by the anisotropic 
$B_2$ term, as described below. 
The quartic $C_2$ term, isotropic, provides stability. 
Since $e_1$ is not the primary order parameter, the term ${1\over2}A_1e_1^2$ 
is often omitted\cite{salje87,blackjphy99}. 
We show below however that this term plays a crucial role in determining the 
morphology; 
very different structures are found for large {\it vs}. small $A_1$. 
The $d_2$ term, which penalizes spatial inhomogeneities in the primary order 
parameter, is necessary to give the domain walls an energy and so prevent 
subdivision into arbitrarily fine variants.  A discussion of other possible
derivative invariants is given in the Appendix.

The H-O transition occurs at $A_2={2\over9}B_2^2/C_2$; 
it is first-order. 
For $A_2>{1\over4}B_2^2/C_2$, the energy has only the H minimum, with 
$e_1=e_2=e_3=0$; 
this minimum disappears for $A_2<0$. 
For $A_2<{1\over4}B_2^2/C_2$, there are in addition three degenerate O 
minima with $e_1=0$, symmetrically located in the $(e_2,e_3)$ plane: 
\renewcommand{\theequation}{3\alph{equation}}
\setcounter{equation}{0}
\begin{eqnarray} 
e_2= e_{20}   \ &,& \  e_3=0                \\
e_2=-e_{20}/2 \ &,& \  e_3=-\sqrt{3}e_{20}/2\\
e_2=-e_{20}/2 \ &,& \  e_3=+\sqrt{3}e_{20}/2  
\end{eqnarray} 
\renewcommand{\theequation}{\arabic{equation}}
\setcounter{equation}{3}
where $
e_{20}=\left[B_2+\left(B_2^2-4A_2C_2\right)^{1/2}\right]
/\left(2C_2\right)\ $; 
at $T_c$, $e_{20}={2\over3}B_2/C_2$. 
The three variants are elongated in the $\hat1$, 
$-{1\over2}\hat1+{\sqrt{3}\over2}\hat2$, and 
$-{1\over2}\hat1-{\sqrt{3}\over2}\hat2$ directions respectively.

\section{variant-variant walls}

We seek the solution linking the two variants of Eqs. (3b) and (3c), the 
results for this pair being simpler than for the others; 
the analysis is easily extended to treat a band of walls. 
The key new results are the expressions (7) for $e_1$ and $\omega_3$ in terms 
of $e_2$ and $e_3$. 

%As for the C-T twin wall\cite{curnoe00}, 
We start from the first-order compatibility relations $u_{i,12}=u_{i,21}$; 
with the linearized strain tensor, these are 
\begin{equation}
\begin{array}{rl}
\partial_2\left(e_1+e_2\right) & = \partial_1\left(e_3+\omega_3\right) \ ,\\
\partial_1\left(e_1-e_2\right) & = \partial_2\left(e_3-\omega_3\right) \ ,
\end{array}
\end{equation}
where $\omega_3 =(u_{1,2}-u_{2,1})/2$ is the $z$ component of the local 
rotation {\boldmath$\omega$}. 
Elimination of $\omega_3$ gives the more familiar second-order relation 
\begin{equation}
 \left(\partial_1^2+\partial_2^2\right) e_1
-\left(\partial_1^2-\partial_2^2\right) e_2
=2\partial_1\partial_2 e_3\ .
\end{equation}
On applying the boundary conditions 
\renewcommand{\theequation}{6\alph{equation}}
\setcounter{equation}{0}
\begin{eqnarray}
e_2(\pm\infty)=-e_{20}/2 \ \ &,& \ \ e_3(\pm\infty)=\pm\sqrt{3}e_{20}/2 \ , \\  
e_1(\pm\infty)=0         \ \ &,& \ \ \omega_3(\pm\infty)=\pm\Omega\ , 
\end{eqnarray}
\renewcommand{\theequation}{\arabic{equation}}
\setcounter{equation}{6}
and demanding that the strains and $\omega_3$ be functions of the single 
variable $X=x_1 \cos\beta +x_2 \sin\beta$, one finds easily the two 
solutions 
\renewcommand{\theequation}{7\alph{equation}}
\setcounter{equation}{0}
\begin{eqnarray}
\beta=0     \ ,\ X=x_1 \ ,\ e_1&=&+(e_2+e_{20}/2)\ ,\ \omega_3=-e_3 \\
\beta=\pi/2 \ ,\ X=x_2 \ ,\ e_1&=&-(e_2+e_{20}/2)\ ,\ \omega_3=+e_3 
\end{eqnarray}
\renewcommand{\theequation}{\arabic{equation}}
\setcounter{equation}{7}
Then, with the other two pairings, one has a total of six possible wall 
orientations, at multiples of $\pi/6$ to the hexagonal 100 
axis\cite{sapriel75}. 
These results are independent of the details of the free-energy density; 
we made only three assumptions: the strains are small, they (and $\omega_3$) 
are functions of a single variable, and they have the asymptotic values (6). 

The two kinds of wall are orthogonal, as already known\cite{sapriel75}. 
They are obviously different in nature; 
it is enough that the angles between the stretch axes are different, 
but we note also that the dilatational strains have opposite sign in the 
wall region. 
The walls need not possess the twin property; 
in Mg-Cd alloys\cite{sinclair77,kitano88}, the ``reflection'' walls are twin 
walls but the ``rotation'' walls are not. 

The rotations in Equations (7) are the key to understanding the domain 
patterns. 

\noindent For $\beta=0$, $X=x_1$: the variant at $x_1\to+\infty$, 
with $e_3>0$, is rotated negatively ($\omega_3<0$, clockwise); 
the other variant is rotated positively; the angle between the stretch 
directions is not $\pi/3$ but rather $\pi/3-\sqrt{3} e_{20}$. 

\noindent For $\beta=\pi/2$, $X=x_2$: the variant at $x_2\to+\infty$, 
with $e_3>0$, is rotated positively (counterclockwise); $\ldots$ ; 
the angle between the stretch directions is not $2\pi/3$ but rather 
$2\pi/3-\sqrt{3} e_{20}$. 

\noindent These rotations are well known in the metallurgical literature, 
where they are obtained from geometrical considerations.
In Mg-Cd alloys, the angles between the stretch axes of the linked variants 
are calculated from the lattice constants to be\cite{sinclair77} 
$57.5^\circ$ or $117.4^\circ$, in good agreement ($\pm\sim1^\circ$) with the 
observations\cite{sinclair77}. 
The leading corrections to Eqs. (7) (due to nonlinear terms in the strain 
tensor) are easily found: 
$e_1     =\pm     \bigl(e_2 + {1\over2}e_{20}\bigr)+{\rm O}\left(e^3\right)$ 
and 
$\omega_3=\mp e_3 \bigl( 1 \mp{1\over2}e_{20}\bigr)+{\rm O}\left(e^3\right)$. 

Our results for H-O walls are in some respects the same as for 
C-T\cite{bk84,curnoe00} and T-O\cite{jacobs85,jacobs00} systems. 
In all three systems, a given pair of variants can be linked 
by one of two possible walls, at $\pi/2$ to each other\cite{sapriel75}. 
A given wall not only links the two variants, but also rotates them 
differentially by an amount proportional to the order parameter, as seen for 
example in Eqs. (7). 
Collisions of wall bands generate new kinds of walls\cite{jacobs00}, including 
a strange ``wall'' between two variants with identical strains but opposite 
rotations. 
In both C-T and T-O systems, where the angle between the variant 
axes is $\pi/2$ minus a correction, the two possible wall structures are 
identical with respect to the dilatational and shear strains; 
in fact, they are related by a symmetry operation, and so have the same energy. 
In H-O systems on the other hand, there is no symmetry operation 
relating the two possible structures and so they generally have different 
energies. 

To determine the structure of the wall linking the variants of Eqs. (3b) and 
(3c), we first use Eqs. (7) to eliminate $e_1$ from Eq. (2) for the density: 
\begin{eqnarray} 
{\cal F}_{wall}=
& & \frac{A_1}{2}\left(e_2+\frac{e_{20}}{2}\right)^2
   +\frac{A_2}{2}\left(e_2^2+e_3^2         \right)
   -\frac{B_2}{3}\left(e_2^3-3e_2e_3^2     \right)    \nonumber\\
& &+\frac{C_2}{4}\left(e_2^2+e_3^2         \right)^2
   +\frac{D_2}{2}\left(\frac{de_2}{dX}     \right)^2
   +\frac{D_3}{2}\left(\frac{de_3}{dX}     \right)^2\ .
\end{eqnarray} 
Here $D_2=D_3=d_2$ for the density of Eq. (2), but $D_2$ becomes 
$D_2=d_1+d_2$ if the $d_1$ term is included in the density (see the Appendix); 
then the length scales of $e_2$ and $e_3$ for the wall can in principle be 
different, even though their gradients appear with the same coefficient. 
Obviously the two kinds of wall ($X=x_1$ and $X=x_2$) have the same energy. 
This degeneracy is however due to the particular form of Eq. (2); 
it is broken if we add to ${\cal F}$ any invariant odd in $e_1$, for example 
the term\cite{yamada89,jacobs95,onuki99} $E e_1(e_2^2+e_3^2)$. 
In principle, such terms can give one wall type a much larger energy than the 
other; 
if so, then the domain patterns may be quite unlike those in cordierite, 
Ta$_4$N, Mg-Cd and lead orthovanadate. 

The compatibility relations are satisfied in Eq. (8), and so the 
Euler-Lagrange equations can be obtained by varying $e_2$ and $e_3$:  
\renewcommand{\theequation}{9\alph{equation}}
\setcounter{equation}{0}
\begin{eqnarray}
&& A_1(e_2+e_{20}/2)+A_2e_2-B_2(e_2^2-e_3^2)+C_2e_2(e_2^2+e_3^2) \nonumber \\
 && \hspace{.5in}  =D_2d^2e_2/dX^2 \ ,  \\
&&A_2e_3+2B_2e_2e_3+C_2e_3(e_2^2+e_3^2)=D_3d^2e_3/dX^2\ , 
\end{eqnarray}
\renewcommand{\theequation}{\arabic{equation}}
\setcounter{equation}{9}
with the boundary conditions of Eq. (6a); 
the $A_1$ term in Eq. (9a) results from the compatibility relations. 
Another route to Eqs. (9) starts from requiring $F$, with the density 
of Eq. (2), to be stationary with respect to the displacement {\bf u}; 
then the strains are assumed to depend only on $x_1$ or $x_2$ and Eqs. (7) 
are used to eliminate $e_1$; 
finally, the results are integrated to give Eqs. (9). 

Equations identical in appearance to Eqs. (9) were derived and solved in our 
study\cite{curnoe00} of C-T ferroelastics. 
At $A_2=-2B_2^2/C_2$, they have an analytical solution\cite{bk84} with 
$e_2=-e_{20}/2$ and so $e_1=0$; 
at any other $T$, however, the dilatational strain is not zero in the wall 
region. 
The analytical solution at $A_2=-2B_2^2/C_2$, numerical solutions at other 
$T$, and numerical and analytical results for the wall energy are given in 
Ref.\onlinecite{curnoe00}. 
These C-T results carry over to H-O systems with only changes in notation; 
we point out however that they seem to differ significantly from the 
results of Ref.\onlinecite{blackjphy99}. 

\section{static structures} 

To reduce the number of unknown parameters, we transform variables by 
\renewcommand{\theequation}{10\alph{equation}}
\setcounter{equation}{0}
\begin{eqnarray} 
e_j     &\to& [2\times10^3B_2/(3C_2)]\,e_j\ ,\\
x_i     &\to& \sqrt{9d_2C_2/(2B_2^2)}\,x_i\ ,\\
{\cal F}&\to& [8\times10^6B_2^4/(81C_2^3)]\,{\cal F}\ .
\end{eqnarray} 
\renewcommand{\theequation}{\arabic{equation}}
\setcounter{equation}{10}
Because we work with the displacement, the first (strain) scaling assumes small 
strains (so that the quadratic terms in the strain tensor can be neglected). 
We also define a dimensionless temperature by $\tau=(T-T_0)/(T_c-T_0)$, and a 
dimensionless anisotropy parameter by 
$\zeta=A_1/A_2(T=T_c)= 9A_1C_2/(2B_2^2)$. 
In terms of the new variables, the density is 
\begin{eqnarray} 
{\cal F}=& & 
 \frac{\zeta}{2}      e_1^2
+\frac{\tau}{2}\left(e_2^2+e_3^2\right)
-\frac{b}{3}\left(e_2^3-3e_2e_3^2\right)             \nonumber\\
&+&\frac{c}{4}\left(e_2^2+e_3^2\right)^2 
  +\frac{1}{2}\left[\left(\vec\nabla e_2\right)^2
                   +\left(\vec\nabla e_3\right)^2\right] \ ,
\end{eqnarray} 
where $b=3\times10^3$ and $c=2\times10^6$; 
the scaled theory then has only two parameters, $\tau$ and $\zeta$. 
The transition ($T=T_c$) is at $\tau=1$, the order parameter at $T_c$ is 
$e_{20}=10^{-3}$ (an arbitrary value) and the analytical solution\cite{bk84} 
is found at $\tau=-9$. 
To relate $\tau$ to the physical temperature $T$, we require both $T_c$ 
and $T_0$; 
the latter can be estimated from the $T$ dependence of the order parameter 
(for example, the value at $T_0$ is 3/2 the value at $T_c$). 

Wishing to study wall configurations, we forced walls into the systems by 
applying periodic boundary conditions to the displacement {\bf u}; 
this apparently simple requirement is sufficient to generate many metastable 
states. 
Free boundary conditions\cite{reid97}, it seems, cannot generate the complex 
structures seen in experiment, and fixed boundary conditions (specifically 
${\bf u}=0$ on and outside the boundary\cite{jacobs95,jacobs00}) give results 
which suffer severely from finite-size effects. 
The static structures described in this section were found by minimizing 
the free energy with respect to the components of {\bf u}, using a 
conjugate-gradient method; 
the iteration was continued until numerical errors prevented further decrease 
in the energy. 
A variety of starting configurations was used, often random displacements. 
Finite-size effects are certainly present in the systems we can study (they 
favour simpler structures); 
we see the basic structures observed experimentally, however, and so 
these effects seem unimportant qualitatively. 

Detailed comparison with experiment is unfortunately not possible for any of 
the materials; 
in fact, experiment offers no guidance at all regarding the parameters 
$\tau$ and $\zeta$ of even the minimal theory. 
Accordingly, we investigated several sets of parameter values; 
the desired comparison can be made when the parameters are measured. 

As the temperature decreases, the order parameter ($e_{20}$) of the 
homogeneous (O) product phase increases and the wall width (proportional to 
$1/e_{20}$) decreases.  
The two curvatures at the O minima also change, but in different ways. 
These are $\gamma_{\parallel}$ and $\gamma_{\perp}$ in the directions 
parallel and transverse to the line from the origin to the O minima: 
\renewcommand{\theequation}{12\alph{equation}}
\setcounter{equation}{0}
\begin{eqnarray} 
\gamma_{\parallel}(T)&=&\tau-2be_{20}+3ce_{20}^2 \ ,\\
\gamma_{\perp}(T)    &=&\tau+2be_{20}+ ce_{20}^2 \ ;
\end{eqnarray} 
\renewcommand{\theequation}{\arabic{equation}}
\setcounter{equation}{12}
both increase monotonically with decreasing $T$. 
For $T_{BK}<T\leq T_c$ ($-9<\tau\leq1$), $\gamma_{\parallel}<\gamma_{\perp}$ 
with $(\gamma_{\parallel},\gamma_{\perp})=(1,9)$ at $T=T_c$. 
At $T=T_{BK}$ ($\tau=-9$), the curvature is isotropic, with 
$\gamma_{\parallel}=\gamma_{\perp}=27$. 
For $T<T_{BK}$, $\gamma_{\parallel}>\gamma_{\perp}$. 
The changing curvatures cause the wall trajectories\cite{curnoe00} 
to start closer to the easy direction in the $(e_2,e_3)$ plane. 

The point of introducing these quantities is that our structures depend 
strongly on the ratio of the curvature $\zeta$ (for dilatational strains) 
to the curvatures $\gamma_{\parallel},\gamma_{\perp}$ (for departures of the 
deviatoric/shear strains from values in a homogeneous O variant). 
This ratio is small for ``soft'' systems and large for ``stiff'' ones. 
This strong dependence on the relative stiffness arises, in the final 
analysis, because the walls rotate the variants as well as link them. 
Because of the rotations, disclinations are formed when walls collide to form 
junctions (or nodes). 
Since a single disclination cannot exist in a crystal, the walls form arrays 
of disclinations of both polarities, often in pairs (dipoles) and higher 
multipoles. 
Finally, soft systems and stiff systems respond very differently to the 
disclinations. 
In soft systems, the walls can deviate considerably from the optimal 
orientations found above; 
additional dilatational strains are then generated, but these have low 
energy since $A_1$ is relatively small. 
In stiff systems, on the other hand, the walls are straight on the whole, 
and their directions are highly constrained; 
other means must be found to minimize the energy. 

To describe the nodes, we use the notation $(k,l,\cdots)$ where the components 
$k,l,\cdots$ are successive angles in units of $\pi/6$ (only approximately); 
this notation, though incomplete because it does not distinguish between 
the two types of walls, is adequate for our purposes. 
The 43 possible disclinations, including the star, have been catalogued 
and the values of the disclination angle $\delta$ given for Mg-Cd 
alloys\cite{kitano88}; 
those with smallest $|\delta|$ follow\cite{kitano88}: 

\noindent 1. $\delta=0$ for the (1,3,5,3) 
node\cite{vicens76,sinclair77,kitano88}. 

\noindent 2. $|\delta|=0.243^\circ$ for the star disclination (a 
three-pointed star with tip angle $\sim\pi/6$, often self-similarly nested), 
the (1,1,1,1,1,7) fan node, and the untitled (1,1,1,4,1,4) node. 

\noindent 3. $|\delta|=0.486^\circ$ for the ``starburst'', our term for a node 
with 12 walls separated by $\sim\pi/6$. 

\noindent 4. $|\delta|=2.472^\circ$ or $3.201^\circ$ for the 
(1,1,1,1,1,1,1,1,4) node. 

\noindent 5. $|\delta|=2.715^\circ$ or $2.958^\circ$ for the (2,5,5), 
(1,1,10) and (1,4,7) three-fold nodes and for two five-fold nodes. 

\noindent Many of the 43 disclinations are not seen because $|\delta|$ is 
too large; 
for example, the at-first-thought obvious (4,4,4) node fails 
by\cite{sinclair77,kitano88} $\sim8^\circ$. 
And one expects nodes with many arms to appear less frequently. 

%number 144 
Figure 1(a) shows a portion of a structure for a soft system ($\zeta=10$) at 
$\tau=-50$, where $(\gamma_{\parallel},\gamma_{\perp}) \approx(117,52)$. 
In this and other structures we found for the same parameter values, we see 
only three-fold nodes and bends, even though (1,3,5,3), star, fan and other 
nodes have smaller disclination angles. 
In such soft systems, the walls can deviate markedly from the high-symmetry 
directions; 
they bend gently on the whole, but sharp bends can occur, especially 
at necks like that at the bottom of the figure. 
Examination of the rotation $\omega_3$ shows that the necking 
results from the collision of identical variants with different orientations. 
As in T-O systems\cite{jacobs00}, the necking reduces the length of the 
collision region where large dilatational strains are generated. 
At $\zeta=1000$, the neck and the $\pi/2$ bends were replaced by a (1,3,5,3) 
node, though with other significant changes. 
Figure 1(a) shows also two well formed (2,5,5) nodes (prominent in 
Ref.\onlinecite{sinclair77}); 
not shown are poorly formed (1,1,10) and (1,4,7) nodes (also seen in 
Ref.\onlinecite{sinclair77}). 
The overall morphology is like that seen in 
cordierite\cite{venk54,putnis87,muller91,salje93}, but lack of detail in 
the published micrographs prevents a direct comparison; 
there seems to be room here for further experiment. 

%number 135 and number 150 
Figures 1(b) and 1(c) show typical structures obtained also at $\tau=-50$, 
but with the moderately soft value $\zeta=100$. 
Nested and unnested stars appear, and also (1,3,5,3), (2,5,5), (1,1,10) and 
(1,4,7) nodes, though all are imperfect. 
Nested stars like that in part (b) are commonly seen in experiment; 
the unnested star of part (c) is like that at the right of Figure 3(c) of 
Ref.\onlinecite{amelinckx80a}. 
%number 156 

\begin{figure}[ht]
\epsfysize=3.4in
\epsfbox[50 140 570 700]{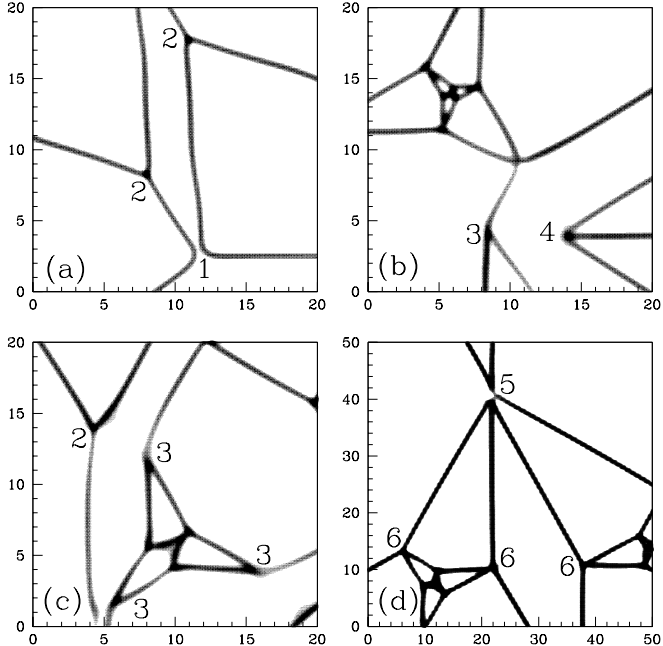}
\caption{
Greyscale plots of $(e_2^2+e_3^2)^{1/2}$ for selected regions of static 
structures. 
In the white regions, the deviatoric strains are near one of the three 
orthorhombic minima; 
the black regions are domain walls. 
Positions 2, 3, and 4 show (2,5,5), (1,4,7) and  (1,1,10)
nodes respectively. 
Part (a) shows a soft system ($\zeta =10$) at low temperature ($\tau=-50$). 
The neck at position 1 separates two domains with the same strains but 
different orientations. 
Parts (b) and (c), both moderately soft systems ($\zeta = 100$) at $\tau=-50$, 
show nested and unnested stars respectively. 
Part (d) shows a moderately soft system ($\zeta = 100$) at a higher 
temperature ($\tau  =-20 $); the length scale is different. 
(1,1,1,4,1,4) and (1,3,5,3) nodes appear at positions 5 and 6. 
}
\end{figure}

Figure 1(d) shows a (1,1,1,4,1,4) node, though the vertex in imperfect. 
This node, similarly imperfect (walls jog at the node), appears 
in Fig. 7(c) of Ref.\onlinecite{vicens76}
and in Figs. 3(b) and 3(c) of Ref.\onlinecite{amelinckx80a}. 
We find that the node tends to split into simpler nodes at lower $T$, 
as the effective stiffness $\zeta/\gamma$ decreases. 
We often observed, but do not display, staircase structures and the ``spiral 
branch configuration'' shown in Fig. 3(a) of Ref.\onlinecite{amelinckx80a}. 
 
%number 156 has (1,1,1,4,1,4) node: tau=0, zeta=50, h=1.0, L=200x1, 
%	d1=10^-6, d2=1
%number 125 
The structure of Figure 2(a), for a stiff system ($\zeta=1000$) at 
$\tau=-50$, shows mostly objects with small disclination angles, 
namely (1,3,5,3) nodes and stars; 
the few three-fold nodes are badly distorted (1,4,7) nodes.  

%number 142 
Figure 2(b) shows a structure formed close to $T_c$. 
The most important features are the triangular and trapezoidal pockets 
of the high-$T$ (hexagonal) phase; 
the first is seen in Refs.\onlinecite{amelinckx80a,amelinckx80b}, 
and the second in Fig. 1(e) of Ref.\onlinecite{amelinckx80b}. 
Trapped high-$T$ phase was not seen by Wen {\it et al}.\cite{wen99} 
who looked at a single temperature, perhaps too low. 
The pockets can be viewed as small grains trapped by rigid walls which 
prevent them from changing shape; 
they can transform to the low-$T$ phase only through formation of internal 
walls, which however cost too much energy at this $T$. 
The structure of Figure 2(b) was generated by warming. 
In the initial structure, the stars were nested and the trapezoids also 
showed internal structure; 
both internal structures gradually faded away with warming.
Figure 2 of Ref.\onlinecite{amelinckx80b} shows comparable 
before-and-after pictures found on cooling. 

Figure 2(b) shows also the fan node, though the vertex is imperfect. 
Imperfect fan nodes are seen 
in Fig. 7(b) of Ref.\onlinecite{vicens76}, 
in Fig. 11(b) of Ref.\onlinecite{sinclair77} (the high-resolution image 
shows 3 three-fold junctions), 
and in Fig. 3(d) of Ref.\onlinecite{amelinckx80a}. 

We found structures also for $\zeta=0.5$, 5 and 50, at $\tau=0$, where 
$(\gamma_{\parallel},\gamma_{\perp})=(4.5,13.5)$; 
the patterns differed significantly in some respects from those described 
above. 
These results and those for $\tau=-50$ suggest that the stiffness ratio 
$\zeta/\gamma$ is of order 1 for Mg-Cd at the temperature of the observations. 

Since $\zeta/\gamma$ depends on $T$, significant structural changes may 
result from changing $T$: 
if $\zeta/\gamma$ is large at $T_c$ and small at low $T$, then stars and other 
structures characteristic of stiff materials may decay as $T$ decreases. 

\section{dynamics} 

We derive the equations of motion using the Lagrangian formulation, starting 
from ${\cal L}={\cal T}-{\cal F}$ where ${\cal T}= \rho (\dot{u}_i)^2/2$ 
is the kinetic-energy density\cite{landau,reid97}.   
The result is 
\begin{equation}
\rho \ddot{u}_i  = \sigma_{ij,j} + \sigma_{ij,j}'
\label{motion1}
\end{equation}
where $\sigma_{ij}=\partial{\cal L}/\partial u_{i,j}$ is the stress tensor
and $\sigma_{ij}'$ is the dissipative stress tensor. 
The most general form for the latter, which describes non-conservative 
forces (friction) in the system, is $\sigma_{ij}'= A_{ijkl}'\dot{u}_{k,l}$ 
to lowest order; 
we note that the dissipation cannot depend on the velocity field $\dot{u}_k$ 
itself, only on its space derivatives. 
The dissipative stress tensor is conveniently found from the dissipative 
function $\Psi$ by $\sigma_{ij}' = \partial \Psi /\partial \dot{u}_{i,j}$. 
The form which respects the symmetry of the parent (hexagonal) phase is
\begin{equation}
\Psi =  \frac{A_1'}{2} \dot{e}_1^2 + \frac{A_2'}{2}(\dot{e}_2^2+
\dot{e}_3^2)\ .
\end{equation}

We assume that the dissipation is large and drop the inertial term 
$\rho \ddot{u}_i$; 
the equations of motion are then 
\begin{eqnarray}
A_1'\dot{e}_{1,1}+A_2'(\dot{e}_{2,1}+\dot{e}_{3,2}) 
& = & -(G_{1,1}+G_{2,1}+G_{3,2})\ , 
\nonumber \\
A_1'\dot{e}_{1,2}-A_2'(\dot{e}_{2,2}-\dot{e}_{3,1}) 
& = & -(G_{1,2}-G_{2,2}+G_{3,1})\ ,
\end{eqnarray}
where $G_i = \delta F/\delta e_i$: 
%G_1 &=& \frac{\delta F}{\delta e_1}  = A_1e_1\\
%G_2 &=& \frac{\delta F}{\delta e_2}  \nonumber \\
%&= & A_2e_2-B_2(e_2^2-e_3^2)
%+C_2(e_2^3+e_3^2e_2) -d_2\nabla^2e_2\\
%G_3 &=&\frac{\delta F}{\delta e_3}  \nonumber \\
%& =& A_2e_3+2B_2e_2e_3
%+C_2(e_2^2+e_3^2)e_3  -d_2\nabla^2e_3
\begin{eqnarray}
G_1 &=& A_1e_1,\\\
G_2 &=& A_2e_2- B_2(e_2^2-e_3^2) +C_2e_2(e_2^2+e_3^2) -d_2\nabla^2e_2,\\
G_3 &=& A_2e_3+2B_2e_2e_3        +C_2e_3(e_2^2+e_3^2) -d_2\nabla^2e_3.
\end{eqnarray}
Actually, we have to use the equations of motion in terms of the 
displacement components: 
\begin{eqnarray}
& & \left(\begin{array}{cc}
A_1'\partial_1^2+A_2'(\partial_1^2+\partial_2^2) & 
A_1'\partial_1\partial_2 \\
A_1'\partial_1\partial_2 & 
A_1'\partial_2^2+A_2'(\partial_1^2+\partial_2^2)  
\end{array} \right) 
\left( \begin{array}{c}
\dot{u}_1 \\ \dot{u}_2 \end{array} \right) \nonumber \\
&& \hspace{.5in} = 
-2\left( \begin{array}{c}
G_{1,1}+G_{2,1}+G_{3,2} \\
G_{1,2}-G_{2,2}+G_{3,1} \end{array} \right) 
\end{eqnarray}
The time is scaled so that $A_2^\prime=1$; 
the dependence on the parameter $A_1^\prime$ was found to be weak during
the H-O transition, but the final states (completely relaxed configurations) 
did depend on $A_1'$.   We used $A_1'=0$ for the results shown in Figure 3.

These equations must be solved numerically. 
Each time step started with the displacement ${\bf u}(x_1,x_2)$ known 
at the points of a $256^2$ grid. 
The space derivatives of $u_i$ were approximated by finite-difference 
expressions centered on a $5\times5$ grid.
The non-linear terms on the right-hand sides were computed, and then the 
equations were fast-Fourier transformed to give algebraic equations. 
The solutions were advanced in time using the Euler method, and then 
fast-Fourier transformed back to position space, to start the next step. 

All our dynamical studies were started by perturbing the high-temperature 
state at a single point. 
The initial state certainly does not represent the nucleus of the 
transformation, but we expect its growth to resemble that in experiment; 
in partial confirmation, we see structures much like those observed. 
The perturbation grows quickly with time, but eventually it collides with 
self-images generated by the boundary conditions; 
of course the collision stage also has no relation to experiment. 
The hexagonal (high-$T$) phase then disappears (unless the temperature is 
close to $T_c$); 
thereafter, we expect our simulations once again to resemble experiment 
(again we see structures much like those observed). 
The relaxation stage which follows is quite slow. 
We observe the formation of the objects already seen in the static studies 
reported above, but more interestingly also well-defined structures 
(streamers, striations) seen infrequently or not at all statically, most of 
which eventually decay. 

In stiff systems (large $\zeta/\gamma$), large domains can be penetrated by 
streamers (thin, wedge-shaped intrusions with relatively thick walls off the 
optimal directions); 
one often sees also striated patterns of two alternating variants. 
The striations and streamers are relics of  the H-O transformation. 
Dilatational stresses occurring at the interface between the high-$T$ and 
low-$T$ phase are relieved by local fine-scale twinning of the low-$T$ 
phase\cite{kohn92}. 
The result is more complicated domain patterns (striations and streamers) 
shortly after the transformation is complete. 
Star embryos appear in the vicinity of triple junctions; 
these form near mis-oriented or curved domain walls after the surrounding 
area has transformed.
In contrast, systems with small $\zeta/\gamma$ need not reduce dilatational 
stresses so strongly, and so show no interesting microstructure. 

The key difference between the dynamics of ferroelastics and the dynamics 
of conventional systems is that the former is nonlocal. 
For example, the retraction of a streamer (or the vanishing of a domain) 
cannot be effected by local changes; 
instead, a global change in the displacement field is required. 
The nonlocality is seen explicitly in theories\cite{kerr,shenoy} where 
effective long-range interactions between the strains are generated on 
eliminating the displacement. 
And the relaxation stage is different also in that the changes from one 
snapshot to the next cannot be predicted by inspection. 

Figures 3(a), 3(b) and 3(c) show expanding regions of  low-$T$ phase, for  
stiffness parameter $\zeta=10$, 100 and  1000 respectively. 
The 12-arm burst of part (a) is, we suggest, best viewed as a starburst 
disclination expanding into the high-$T$ phase; 
it resembles strongly the ``stellate twin'' of Fig. 1 of 
Ref.\onlinecite{venk54} (though the latter is fully developed), consistent 
with our identification of cordierite as a soft system. 
The stiffer systems of parts (b) and (c) evolve more slowly; 
they also show branching at the outer edges, as discussed above. 

Figure 3(d) shows the system of 3(b) at a later time. 
There are several streamers, which retract very slowly with time; 
similar structures are seen in Fig. 14(b) of Ref.\onlinecite{sinclair77}.
The long timescale associated with this simple evolution (there is no change
in the topology of the domain walls) results from the nonlocal relaxation. 

Figures 3(e) and 3(f) show the system in 3(c) at a later time.  
The complicated structures in 3(e) are not observed in softer systems.  
In stiff systems they are transitory, existing only immediately after 
completion of the phase transformation. 
The domain walls near the striations and star embryos are relatively 
broad and they deviate from their optimal orientations; 
at later times, for the most part, they are replaced by well-defined walls 
as shown in the fully relaxed configuration of 3(f). 
The divided diamond-shaped structure seen at the top of Figure 3(f) 
is seen in Fig. 9 (especially part (c)) of Ref.\onlinecite{vicens76}, 
suggesting that Ta-N is a stiff system.

\section{summary}
A Landau theory of proper hexagonal-orthorhombic ferroelastics, with the 
fewest possible number of parameters, describes qualitatively many of the 
structures seen in Ta$_4$N, Mg-Cd alloys and Mg-cordierite, and in 
trigonal-monoclinic lead orthovanadate. 
Our static and dynamical studies found structures not previously seen 
theoretically, including trapped high-temperature phase, 12-armed bursts, 
streamers and striations. 

Since the theoretical understanding of structures in these systems seems 
now well advanced, we suggest that the next major theoretical task is to 
include noise and fluctuation effects in the time development. 
The effect of the inertial term also seems worth investigation. 

\section*{Appendix} 
Equation (2) contains only one term quadratic in the strain gradients; 
the most general such expression includes four additional invariants: 
\begin{eqnarray}
& & {1\over2}d_1 \left(\vec\nabla e_1\right)^2 \nonumber \\
&+& d_3 
   \left[e_{1,1}\left(e_{2,1}+e_{3,2}\right)+e_{1,2}\left(-e_{2,2}-e_{3,1}
\right)\right] \nonumber \\
&+& d_4\left[\left(e_{2,1}-e_{3,2}\right)^2
-\left(e_{2,2}+e_{3,1}\right)^2\right]\nonumber\\
&+&d_5\left(e_{2,1}e_{3,2}-e_{2,2}e_{3,1}\right) 
\end{eqnarray}
The $d_1$ term may be important more generally, but it can be neglected for
the wall structures of interest to us here;
it obviously plays no role for stiff systems where $|e_1|$ is small, and we 
find numerically that it is qualitatively unimportant also in soft systems. 
Given this, and that we want a minimal theory, we omit the $d_3$ term also. 
The $d_5$ term integrates to a surface term, which vanishes for periodic 
boundary conditions. 

The remaining term (coefficient $d_4$) is closely related to the $d_2$ 
term in Equation (2). 
To demonstrate this, we consider the sum 
\begin{eqnarray}
I=& & {1\over2}d_2'\left[e_{3,1}^2 + (e_{3,1}')^2 +(e_{3,1}'')^2\right]
 \nonumber \\
  &+& {1\over2}d_4'\left[e_{3,2}^2 + (e_{3,2}')^2 + (e_{3,2}'')^2\right]
\end{eqnarray}
of two non-negative invariants. 
The primes represent terms obtained by $2\pi/3$ rotations about 
the hexagonal axis; 
for example, $e_{3,1}^\prime={1\over2}(-\partial_1+\sqrt{3}\partial_2)
{1\over2}(\sqrt{3}e_2-e_3)$. 
%the invariants are non-negative, and so $d_2'>0$ and $d_4'>0$ for stability. 
Each of the above six terms corresponds to the change in the order parameter
in the direction normal to one of the six different domain-wall orientations 
discussed in Section III.  
%  Note that the two terms proportional to $d_2'$ and $d_4'$ are each 
%hexagonal invariants.
The explicit expression for $I$ is 
\begin{eqnarray}
I={3\over16}d_2'& &[3(e_{3,1}^2+e_{2,2}^2)+e_{2,1}^2 + e_{3,2}^2 \nonumber\\
& &+ 2(e_{2,1}e_{3,2}+e_{2,2}e_{3,1}) ] \nonumber \\
+ {3\over16}d_4'& &[e_{3,1}^2+e_{2,2}^2 +3( e_{2,1}^2+e_{3,2}^2)\nonumber \\
& &- 2(e_{2,1}e_{3,2}+e_{2,2}e_{3,1}) ]\ . 
\end{eqnarray}
Equivalent results are found on working with $e_2$ rather than $e_3$, for 
\begin{eqnarray}
  e_{2,i}^2 + (e_{2,i}')^2 +(e_{2,i}'')^2
 =e_{3,j}^2 + (e_{3,j}')^2 +(e_{3,j}'')^2\ ,
\end{eqnarray}
with $i\neq j$. 

The isotropic case corresponds to $d_2'=d_4'$; then $I$ equals the
$d_2$ invariant we used in the free energy.  
The $d_4$ invariant  is found from the difference, $d_2'=-d_4'$.  
In keeping with our goal of a minimal theory, we consider the isotropic case
and omit the $d_4$ term in the free energy. 

\acknowledgments
We thank Colin Borys for technical assistance.
This research was supported by the Natural Sciences and Engineering 
Research Council of Canada.

\pagebreak
\section*{Colour Figures}
\begin{figure}[ht]
\caption{Samples of static configurations. 
The three different orthorhombic variants are represented as red, blue and 
green; 
the high-temperature phase is represented as black.
Part (a) shows a very stiff system ($\zeta = 1000$) at low temperature 
($\tau = -50$); the microstructure is dominated by nested stars and 
(1,3,5,3) zero-disclination nodes. 
Part (b) shows another very stiff system ($\zeta=1000$), but at $\tau=0.8$, 
just below the transition temperature ($\tau = 1$). 
The important features are the triangular and trapezoidal regions of trapped 
high-$T$ phase and the (1,1,1,1,1,7) fan nodes. 
}
\end{figure}

\begin{figure}[ht]
\caption{Snapshots of dynamical simulations at different times for 
identical starting conditions and different values of the stiffness parameter, 
at $\tau=-50$ (well below the transition temperature).  
Each frame has size $51.2\times 51.2$; 
the colour scheme is as in Figure 2. 
Parts (a), (b) and (c), with $\zeta = 10$, 100 and 1000 respectively, show 
early-time structures (t=0.255) evolving from the perturbed high-$T$ phase. 
The transformed region expands and then collides with images of itself 
because of the periodic boundary conditions; 
at later times the transformed phase occupies the whole frame. 
Part (d), a continuation of part (b)  at time $t=8.191$, shows several narrow 
domains or ``streamers".
Parts (e) and (f) are continuations of part (c) at times $t=0.511$ and
$t=16.383$ respectively.  
Part (e) shows transitory structures associated with stiff systems, namely
bi-domain striations and star embryos.  
Part (f) is a fully relaxed configuration.
}
\end{figure}

\pagebreak

\begin{figure}[ht]
\epsfysize=3.15in
\epsfbox[0 280 550 580]{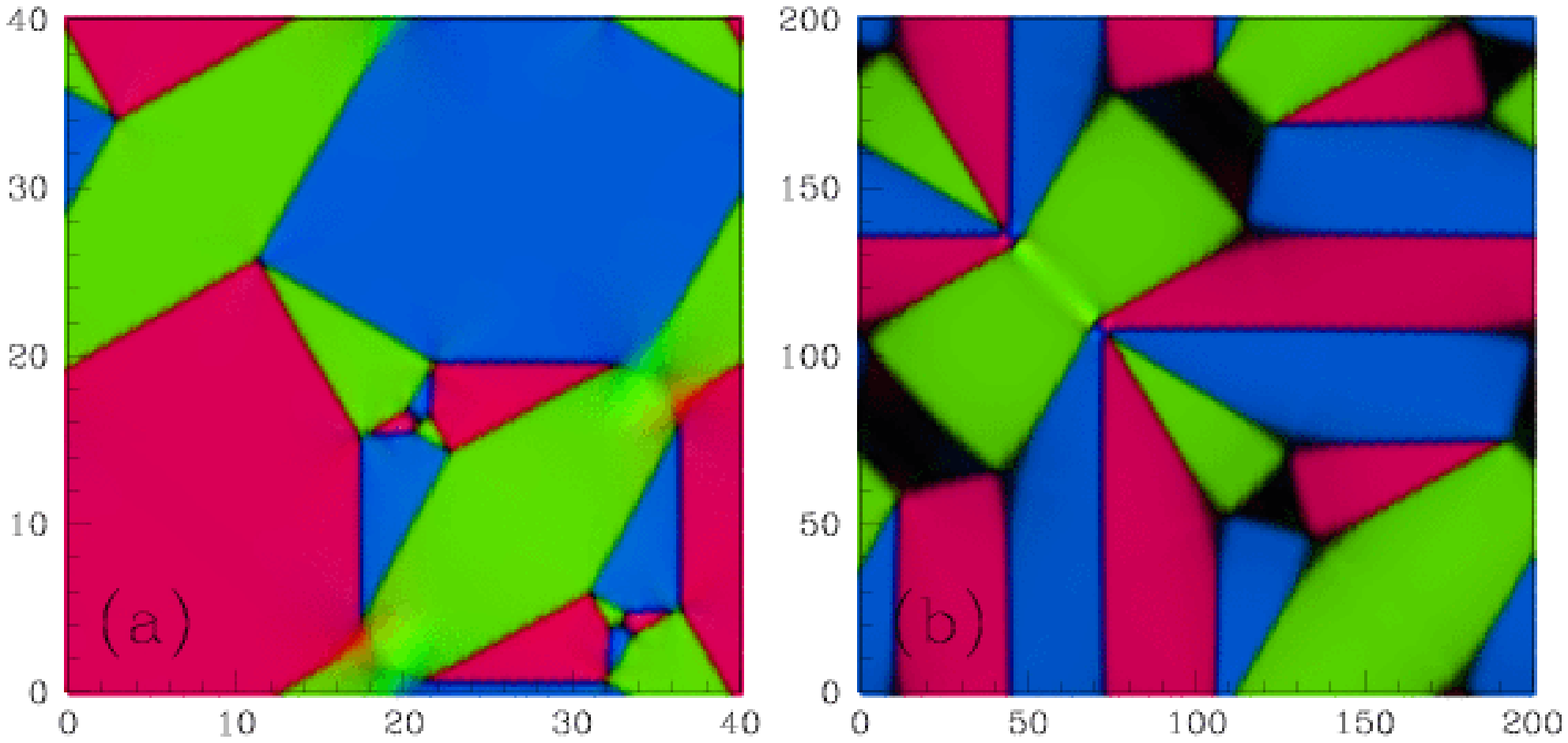}
\end{figure}
Figure 2

\begin{figure}[ht]
\epsfysize=4.20in
\epsfbox[0 200 600 560]{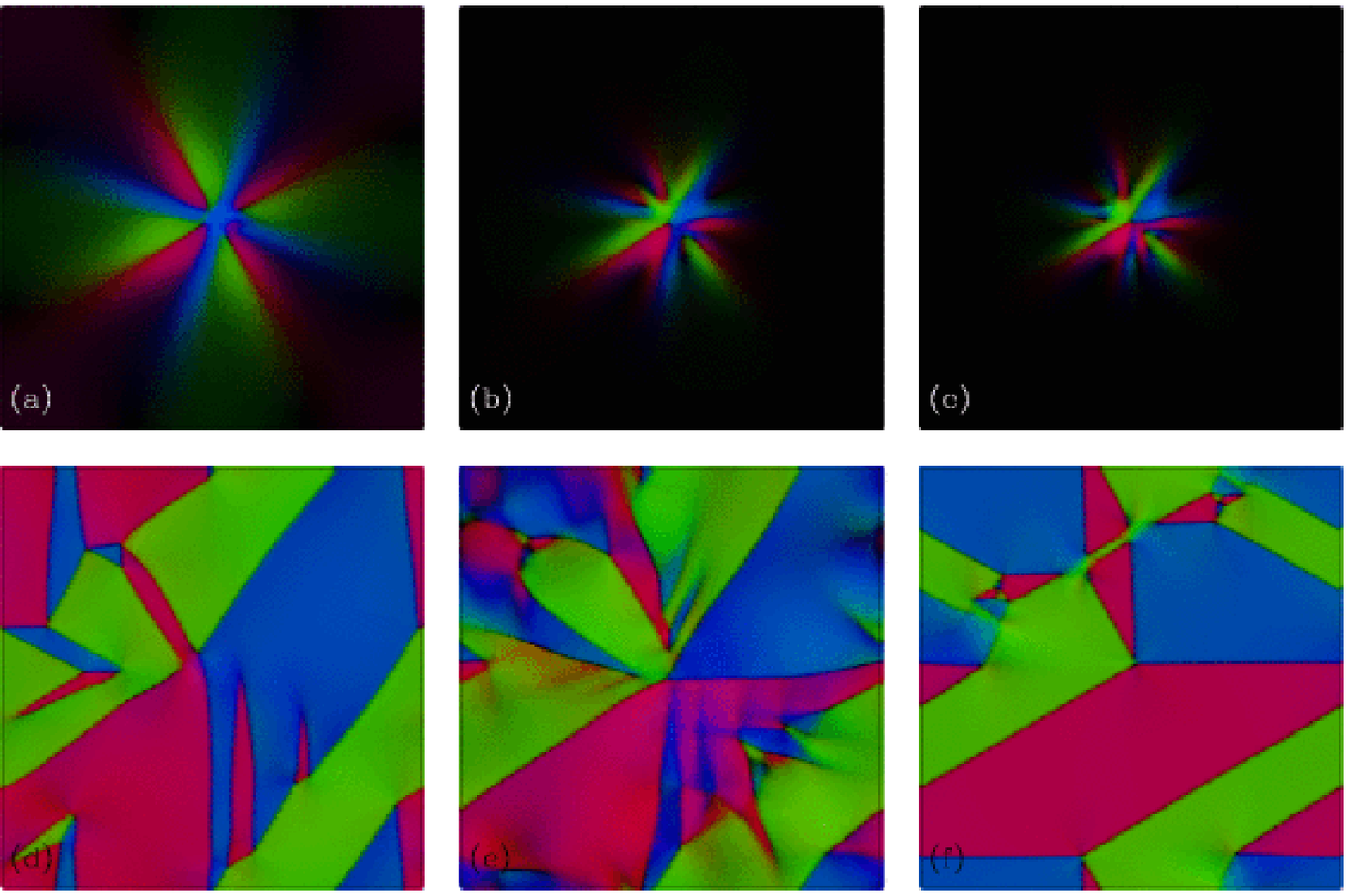}
\end{figure}
Figure 3

\end{document}